\documentclass[%
 reprint,superscriptaddress,
 amsmath,amssymb,prb,
]{revtex4-2}

\usepackage{graphicx}% Include figure files
\usepackage{dcolumn}% Align table columns on decimal point
\usepackage{bm}% bold math
\usepackage{color}
\usepackage{siunitx}
\usepackage[version=3]{mhchem} % Formula subscripts using \ce{}
\newcommand{\tred}[1]{\textcolor{black}{#1}}
\def \ZVPO{\ce{Zn2VO(PO4)2}}

\def \TN{$T_{\mathrm N}$}

\begin{document}

\preprint{APS/123-QED}

\title{Quantum and classical spin dynamics across temperature scales in the $S=1/2$ Heisenberg antiferromagnet} \thanks{This manuscript has been authored by UT-Battelle, LLC under Contract No. DE-AC05-00OR22725 with the U.S. Department of Energy.  The United States Government retains and the publisher, by accepting the article for publication, acknowledges that the United States Government retains a non-exclusive, paid-up, irrevocable, world-wide license to publish or reproduce the published form of this manuscript, or allow others to do so, for United States Government purposes.  The Department of Energy will provide public access to these results of federally sponsored research in accordance with the DOE Public Access Plan (http://energy.gov/downloads/doe-public-access-plan).}

\author{Pyeongjae Park}
\email{parkp@ornl.gov}
\affiliation{Materials Science \& Technology Division, Oak Ridge National Laboratory, Oak Ridge, TN 37831, USA}

\author{G. Sala}
\affiliation{Oak Ridge National Laboratory, Oak Ridge, TN, 37831, USA}

\author{Daniel M. Pajerowski}
\affiliation{Neutron Scattering Division, Oak Ridge National Laboratory, Oak Ridge, Tennessee 37831, USA}

\author{Andrew F. May}
\affiliation{Materials Science \& Technology Division, Oak Ridge National Laboratory, Oak Ridge, TN 37831, USA}

\author{James A. Kolopus}
\affiliation{Materials Science \& Technology Division, Oak Ridge National Laboratory, Oak Ridge, TN 37831, USA}

\author{D. Dahlbom}
\affiliation{Neutron Scattering Division, Oak Ridge National Laboratory, Oak Ridge, Tennessee 37831, USA}

\author{Matthew B. Stone}
\email{stonemb@ornl.gov}
\affiliation{Neutron Scattering Division, Oak Ridge National Laboratory, Oak Ridge, Tennessee 37831, USA}

\author{G\'abor B. Hal\'asz}
\email{halaszg@ornl.gov}
\affiliation{Materials Science \& Technology Division, Oak Ridge National Laboratory, Oak Ridge, TN 37831, USA}

\author{Andrew D. Christianson}
\email{christiansad@ornl.gov}
\affiliation{Materials Science \& Technology Division, Oak Ridge National Laboratory, Oak Ridge, TN 37831, USA}
%\date{\today}% It is always \today, today, but any date may be explicitly specified

\begin{abstract}
Using the framework of semi-classical Landau-Lifshitz dynamics (LLD), we conduct a systematic investigation of the temperature-dependent spin dynamics in the $S=1/2$ Heisenberg square-lattice antiferromagnet (SqAF). By performing inelastic neutron scattering measurements on \ZVPO{} (ZVPO) and corresponding finite-temperature spin dynamics simulations based on LLD, we present a comprehensive analysis that bridges quantum and classical spin dynamics over a broad temperature range. First, a remarkable agreement between experimental data and LLD simulations is found in the paramagnetic phase of ZVPO, demonstrating the capability of LLD in accurately determining the spin Hamiltonian of $S=1/2$ systems and capturing the quantum-to-classical crossover of their spin dynamics. Second, by analyzing the discrepancies between the experimental data and the LLD simulations at lower temperatures, we determine the experimental temperature dependence of the quantum effects in the excitation spectrum of the $S=1/2$ SqAF: the quantum renormalization factor for the magnon energies and the quantum continuum above the one-magnon bands. Notably, the emergence of each quantum effect is found to correlate with the formation of three-dimensional long-range order. This work demonstrates the utility of LLD in gaining experimental insights into the temperature-induced modifications of quantum spin dynamics and their convergence towards classical expectations at higher temperatures. This motivates further applications to more challenging quantum antiferromagnets dominated by stronger quantum fluctuations.

\end{abstract}

\maketitle

Understanding the collective phenomena of quantum antiferromagnets remains an ongoing challenge, especially in low-dimensional systems where quantum fluctuations are enhanced relative to three-dimensional antiferromagnets. Recent inelastic neutron scattering (INS) studies on various two-dimensional (2D) $S=1/2$ antiferromagnets have unveiled a plethora of quantum phenomena which illustrate deviations from classical spin dynamics. These deviations include the significant renormalization of magnon energies \cite{BCSO_Ma,BCSO_INS, BCSO_avoided,Sq_NLSWT1}, pronounced magnon decay \cite{YbCl3_INS,CoI2_nphys}, a highly structured multi-magnon continuum \cite{YbCl3_INS,YbCl3_field}, and most interestingly, the emergence of fractionalized excitations \cite{Mourigal_2015,Christensen_2007, BCSO_INS, TLAF_nearQSL, BLCTO, RuCl3_sci}. While these phenomena highlight the diverse landscape of quantum magnetism, they also introduce hurdles in modeling the excitation spectra with traditional semi-classical methods, such as linear spin-wave theory (LSWT), which are typically used to deduce spin Hamiltonians. Thus, the accurate identification of the spin Hamiltonian in a quantum antiferromagnet frequently calls for either sophisticated quantum spin dynamics calculations \cite{CsYbSe2_field,YbCl3_field, BCSO_spinon} or artificially suppressing inherent quantum fluctuations by, for example, applying a strong external magnetic field \cite{YbCl3_field,BCSO_plateau,RYbSe2_plateau}. Notably, both approaches require considerable theoretical and/or experimental efforts. 

Introducing thermal fluctuations by increasing the temperature---a cornerstone variable in condensed matter physics---can significantly aid in exploring the spin dynamics of quantum antiferromagnets. Raising the temperature moves quantum magnets into a regime where thermal fluctuations begin to dominate quantum fluctuations, which enables the following two investigative approaches. First, for temperatures $T$ substantially exceeding the magnetic energy scale, one can determine the spin Hamiltonian using straightforward semi-classical spin dynamics theories. Second, and perhaps more intriguingly, one can then track deviations from semi-classical spin dynamics as $T$ decreases towards the magnetic energy scale, which reveals how quantum fluctuations begin to impact the behavior of the system. Such a "quantum-to-classical" crossover can potentially deepen our understanding of the collective quantum phenomena found in $S=1/2$ antiferromagnets. This approach, however, requires a simulation technique that systematically incorporates thermal fluctuations into spin dynamics, which has been deemed challenging \cite{rescale_Dahlbom}.

Recent studies have made substantial progress in addressing the aforementioned technical challenge using a range of methodologies \cite{Gd2PdSi3_ORF,Kitaev_Paddison,rescale_Dahlbom}. In particular, employing Landau-Lifshitz dynamics (LLD) to numerically calculate dynamical spin-spin correlations via the time evolution of real-space spin configurations emerges as especially promising \cite{Sunny_ref1,Sunny_ref2}. The primary benefit of this approach lies in its ability to provide an energy-resolved profile of spin dynamics at finite temperatures. This allows for the fitting of multi-dimensional dynamical structure factors $S(\mathbf{q},\omega)$, which, thanks to the rich information available across both momentum and energy dimensions, facilitates the determination of the spin Hamiltonian with high fidelity, despite the typically broad and diffuse spectra found at high temperatures \cite{CoI2_nphys,BLCTO}. Moreover, the recent introduction of a temperature-dependent renormalization technique associated with the size of magnetic moments \cite{rescale_Dahlbom} has vastly extended the temperature range over which the LLD approach can accurately capture the energy scale of magnetic excitations; see the Supplemental Material for more explanation \cite{supp}. While these features make LLD a promising approach for modeling spin dynamics under sizable thermal fluctuations, its predictions for $S=1/2$ quantum antiferromagnets have not yet been comprehensively compared with temperature-dependent experimental data. In contrast, for classical systems with larger $S$ values ($S>1/2$), the interplay between thermal fluctuations and spin dynamics has been investigated by several studies \cite{Rb2MnF4_QCC, MgCr2O4_LLD, rescale_Dahlbom}. 

%%%%%%%%%%%%%%%%%%%%%%% Figure - basic information %%%%%%%%%%%%%%%%%%%%%%%%%%%%%%%%
\begin{figure}[t]
\includegraphics[width=1\columnwidth]{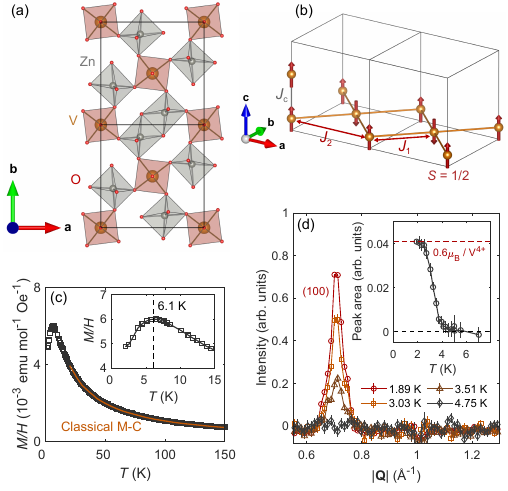} 
\caption{\label{crystal} Structure and magnetic properties of \ZVPO{} (ZVPO). (a) Crystal structure of ZVPO. (b) Magnetic structure of ZVPO \cite{ZVPO_pnd} and exchange interactions between V$^{4+}$ magnetic moments. (c) Temperature-dependent magnetization of ZVPO under a 1 kOe magnetic field. The inset highlights the position of maximum susceptibility ($T_{\mathrm{max}} =$ 6.1\,K = 1.63\TN{}). The orange solid line is the magnetic susceptibility obtained from the classical Monte-Carlo simulation with the exchange parameters from the high-temperature INS data analysis (Fig.~\ref{highT}). (d) Elastic component ($-0.1 \,\mathrm{meV}< E < 0.1 \,\mathrm{meV})$ of the neutron scattering data after subtracting the result measured at 10 K, which shows the temperature dependence of the (100) magnetic Bragg peak. The inset shows the peak area fitted by a Gaussian function.}
\end{figure}
%%%%%%%%%%%%%%%%%%%%%%%%%%%%%%%%%%%%%%%%%%%%%%%%%%%%%%%%%%%%%%%%%%%%%%%%%%%%

The $S=1/2$ Heisenberg square-lattice antiferromagnet (SqAF) stands out as an ideal system for assessing the capabilities and limitations of LLD in describing the spin dynamics of quantum antiferromagnets under sizable thermal fluctuations. This is because the quantum effects observed in its excitation spectrum are well-established by extensive previous studies \cite{SqAF_dyn_ref1, SqAF_dyn_ref2, Mourigal_2015, Christensen_2007, La2CuO4, Sq_NLSWT1}, and are less complex than those in other lattice geometries \cite{BCSO_INS, YbCl3_INS}. Key quantum phenomena in the $S=1/2$ SqAF include (i) a momentum-independent spin-wave energy enhancement due to an overall quantum renormalization factor $Z_{c}$ \cite{Zc_SE,Zc_SWT,Sq_NLSWT1}, (ii) a two-magnon continuum above the one-magnon modes \cite{Sq_NLSWT1, Sq_NLSWT2, Sq_NLSWT3, Christensen_2007}, and (iii) downward renormalization along with continuum scattering around $\mathbf{q}=(\pi, 0)$ attributed to fractionalized quasi-particles \cite{Mourigal_2015, Christensen_2007}. Despite these features, the overall magnetic excitation spectrum remains similar to classical spin dynamics predictions. Hence, a comparative analysis with LLD simulations, even though discrepancies are anticipated at low temperatures, can yield insightful conclusions.

This work presents a successful application of the semi-classical LLD approach to elucidate the temperature-driven evolution of quantum spin dynamics in the $S=1/2$ Heisenberg SqAF. By conducting INS measurements of the $S=1/2$ Heisenberg SqAF \ZVPO{} and corresponding finite-temperature spin dynamics simulations through a standardized protocol proposed in this study, we engage in a thorough comparison of quantum versus classical spin dynamics over a broad temperature range, 0.5\TN{}\,$< T < 21.3$\TN{}, with \TN{} being the N\'eel temperature. At temperatures well above 5\TN{}, LLD simulations correspond remarkably well with experimental observations and yield reliable determination of exchange parameters. Notably, this agreement persists down to approximately 1.1\TN{} upon incorporating an additional scale factor for the excitation energies to account for the quantum renormalization factor $Z_{c}$ not included in the LLD simulation. Moreover, by analyzing the discrepancy between measured spectra and LLD simulations at lower temperatures, we delve into how thermal fluctuations dissipate quantum effects within the excitation spectrum of the $S=1/2$ SqAF. The discussion extends to potential applications for other quantum magnetic systems and possible future improvements to the current LLD framework.
 
%%%%%%%%%%%%%%%%%%%%%%% Figure - high-T data fitting %%%%%%%%%%%%%%%%%%%%%%%%%%%%
\begin{figure*}[ht]
\includegraphics[width=0.9\textwidth]{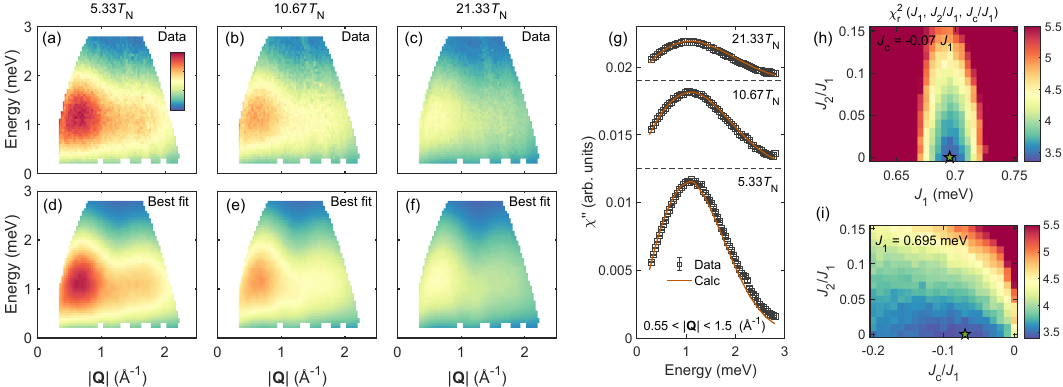} 
\caption{\label{highT} Spin Hamiltonian obtained by analyzing energy- and momentum-resolved excitation spectra in the classical regime ($T \gg$ \TN{}). (a)-(c) Experimental dynamical susceptibility $\chi''(|\mathbf{q}|,E)$ of ZVPO measured at temperatures above \TN{}. \tred{Background signals in a quasielastic region and near the edge of the detector's coverage are masked.} (d)-(f) The best-fitted $\chi''(|\mathbf{q}|,E)$ to (a)-(c) obtained by LLD. (g) Comparison between the measured and simulated energy dependence of momentum-integrated $\chi''(|\mathbf{q}|,E)$. (h)-(i) Slices of the three-dimensional goodness-of-fit ($\chi_{\mathrm{r}}^{2}$, see Supplemental Material \cite{supp}) map around the optimal solution $J_{1}$ = 0.695(15) meV, $J_{2}$ = 0.0(1)$J_{1}$, and $J_{c}$ = -0.07(+7,-20)$J_{1}$ plotted as a green star. The uncertainty of the parameters was derived from the increase in $\chi_{\mathrm{r}}^{2}$ of 1.}
\end{figure*}
%%%%%%%%%%%%%%%%%%%%%%%%%%%%%%%%%%%%%%%%%%%%%%%%%%%%%%%%%%%%%%%%%%%%%%%%%%%%%

The synthesis and characterization methods of polycrystalline ZVPO are provided in the Supplemental Material \cite{supp}. INS data were collected with 9.3\,g of powder ZVPO at the CNCS time-of-flight spectrometer at the Spallation Neutron Source (SNS), using incident neutron energies of 3.32 meV and 1.55 meV. Data were collected at 23 different temperatures between 1.9\,K (base) and 80\,K under standard high-flux chopper conditions. No background subtraction was performed on the data presented in this work. Instead, a constant background was estimated during the comparative analysis with the calculated magnetic excitation spectra for $T\gg$\,\,\TN{} (see the Supplemental Material \cite{supp}). The energy- and momentum-resolved dynamical susceptibility $\chi''(\mathbf{q},E)$ was calculated from the time evolution of the spin system simulated by LLD, using the \textit{Su(n)ny} package \cite{Sunny,Sunny_ref1}. A detailed, step-by-step protocol for this calculation is described in the Supplemental Material \cite{supp}. Instrumental energy and momentum resolutions for the simulation were estimated using the \textit{Pychop} package in Mantid \cite{mantid} and the full width at half-maximum of the (100) magnetic Bragg peak of ZVPO along the $|\mathbf{q}|$ axis, respectively.

ZVPO is a nearly ideal realization of the $S=1/2$ Heisenberg SqAF (Fig.~\ref{crystal}). Previous bulk characterizations and powder neutron diffraction studies have identified the onset of long-range order below the N\'eel temperature \TN{}\,=\,$3.75$\,K \tred{\cite{ZVPO,ZVPO_pnd,ZVPO_Ref2}}. Neutron diffraction data further revealed a N\'eel-type spin configuration for each V$^{4+}$ square-lattice layer, with interlayer alignment being ferromagnetic; see Fig.~\ref{crystal}(b) \cite{ZVPO_pnd}. The magnetic moments of V$^{4+}$ ions are oriented along the $c$-axis. The elastic component ($-0.1 \,\mathrm{meV}< E < 0.1 \,\mathrm{meV}$ for $E_{i}=3.32\,\mathrm{meV}$ neutrons) of our powder INS data reveals a (100) magnetic Bragg peak for $T<3.8$\,K [Fig.~\ref{crystal}(d)], consistent with previous studies \cite{ZVPO_pnd}.

The magnetism of ZVPO can be described by the minimal spin model
\begin{align}\label{Hamiltonian}
\mathcal{\hat H} = \, & J_{1} \sum_{<i,j>_{1}} \hat {\mathbf{S}}_i \cdot \hat {\mathbf{S}}_{j} + J_{2} \sum_{<i,j>_{2}} \hat {\mathbf{S}}_i \cdot \hat {\mathbf{S}}_{j} \nonumber \\ 
+ \, & J_{c} \sum_{<i,j>_{c}} \hat {\mathbf{S}}_i \cdot \hat {\mathbf{S}}_{j},
\end{align}
where $J_{n}$ and $J_{c}$ denote the $n^{\mathrm{th}}$-neighbor intralayer and first-neighbor interlayer coupling strengths, respectively  [see Fig.~\ref{crystal}(b)]. The known magnetic structure of ZVPO dictates that $J_{c}$ must be negative (i.e., ferromagnetic) and $J_{2}$ should be less than $0.4J_{1}$ \cite{balents_2012}. Prior quantitative estimation of the $J_{1}$ and $J_{2}$ parameters,  based on fitting the temperature-dependent magnetization data to high-temperature series expansion (HTSE) predictions, yielded $J_{1}$ = 0.682\,meV and $J_{2}$ = 0.025$J_{1}$ \cite{ZVPO}. Also, the isotropy of the spin Hamiltonian [Eq.~(\ref{Hamiltonian})] is supported by the observation of gapless Goldstone modes in ZVPO, even though a tiny Ising-type XXZ anisotropy is anticipated from the spin configuration parallel to the $c$-axis; see the Supplemental Material \cite{supp}. Thus, ZVPO can be considered a nearly ideal $S=1/2$ nearest-neighbor Heisenberg SqAF, as also confirmed by the analysis presented in this work (Fig.~\ref{highT}).

%%%%%%%%%%%%%%%%%%%% Figure - T-dependent spin-wave analysis %%%%%%%%%%%%%%%%%%%%%%%%%%
\begin{figure*}[ht]
\includegraphics[width=0.95\textwidth]{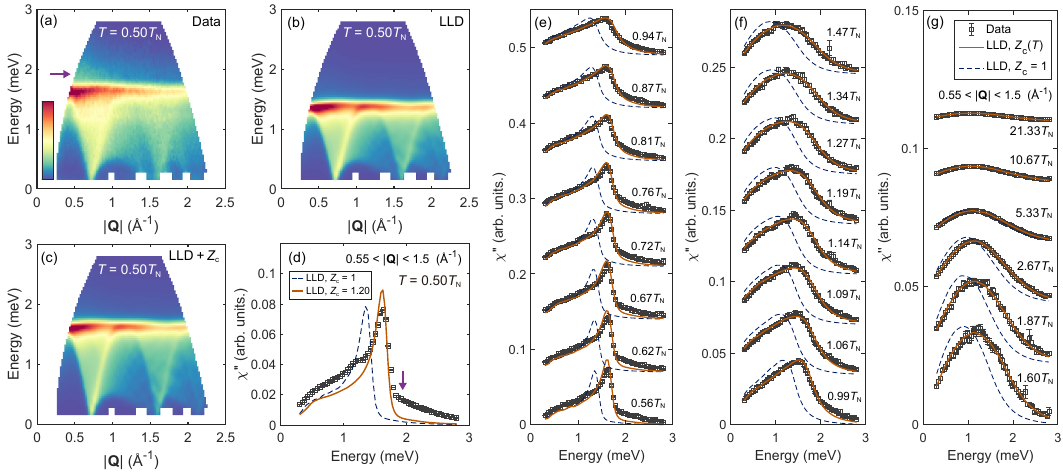} 
\caption{\label{spinwave} Comparison between the INS data and the LLD simulations at low temperatures. (a)-(d) Measured and simulated $\chi''(|\mathbf{q}|,E)$ maps at $T=$\,0.5\TN{}. Compared to the LLD result in (b), the result in (c) further involves the quantum renormalization factor $Z_{c} > 1$, as explained in the main text. \tred{The low-energy signal observed at around 2\AA$^{-1}$ in the data originates from acoustic phonons.} (e)-(g) Full temperature evolution of the magnetic excitation spectrum plotted via the dynamical susceptibility $\chi''(|\mathbf{q}|,E)$, demonstrating the excellence of the LLD approach in capturing the spin dynamics above \TN{} and visualizing the transition from quantum to classical spin dynamics. Orange solid (blue dashed) lines are the LLD simulation results with (without) the quantum renormalization factor $Z_{\mathrm{c}}$.}
\end{figure*}
%%%%%%%%%%%%%%%%%%%%%%%%%%%%%%%%%%%%%%%%%%%%%%%%%%%%%%%%%%%%%%%%%%%%%%%%%%%%%
The investigation of the spin dynamics in the predominantly classical regime ($T\gg$\,\TN{}), when based on the analysis protocol suggested in this work (see the Supplemental Material \cite{supp}), allows for credible estimation of the exchange parameters through LLD (Fig.~\ref{highT}). Notably, simultaneous analysis of the energy-resolved excitation spectra measured at various temperatures provides enough information to determine multiple exchange parameters. The optimal values of $J_{1}$, $J_{2}$, and $J_{c}$ were identified by finding the minimum of the reduced chi-square $\chi_{r}^{2}(J_{1}, J_{2}, J_{c})$ between the three measured [Figs.~\ref{highT}(a)-(c)] and calculated [Figs.~\ref{highT}(a)-(f)] dynamical susceptibility maps $\chi''(|\mathbf{q}|,E)$ \cite{supp}. The agreement between the data and the best fit is exceptional, as evident in the energy-dependent profile of $\chi''(|\mathbf{q}|,E)$ shown in Fig. \ref{highT}(g).

The minimal $\chi_{\mathrm{r}}^{2}$ is found for $J_{1}$ = 0.695(15)\,meV, $J_{2}$ = 0.0(1)$J_{1}$, and $J_{c}$ = -0.07(+7,-20)$J_{1}$, with the uncertainties derived from the increase in $\chi_{\mathrm{r}}^{2}$ of 1. The uncertainty in $J_{c}$ is asymmetric along its positive and negative directions, thus both values are noted. While these results closely align with the $J_{1}$ and $J_{2}$ values derived from the HTSE analysis of thermodynamic measurements \cite{ZVPO}, our analysis indicates more clearly that $J_{2}$ is negligible [Figs. \ref{highT}(h)--(i)]. Also, it points towards a more accurate model that incorporates finite $J_{c}$ ($< 0$), an essential element for understanding \TN{} in the $S=1/2$ SqAF and the ferromagnetic spin configuration along the $c$-axis [Fig. \ref{crystal}(b)]. While the $\chi_{r}^{2}$ metric indicates non-zero $J_{c}$, its relatively minor impact on the excitation spectra in Figs. \ref{highT}(a)-(c) results in a significant uncertainty [Fig. \ref{highT}(i)]. Nevertheless, the obtained $J_{c}$ = -0.07$J_{1}$ is well corroborated by our analysis of the $T\ll$\,\TN{} data represented fully by a spin-wave spectrum, which exhibits a unique feature that enables more precise quantification of $J_{c}$ (see the Supplemental Material \cite{supp}). The credibility of the fitted parameters is also evident in the consistency between the measured and calculated temperature dependence of the magnetic susceptibility [Fig.~\ref{crystal}(c)], the latter derived from classical Monte-Carlo simulations with the spin length normalized to $\sqrt{S(S+1)}$ \cite{supp}. 

Based on the exchange parameters derived in the classical regime ($T\gg$\,\TN{}), we extend our analysis to lower temperatures where sizable quantum effects on the spin dynamics are expected. Specifically, we present a comprehensive overview of the temperature effects based on the measured excitation spectra and LLD simulation results at 20 different temperatures ranging from 0.50\TN{} and 2.67\TN{}. It is important to emphasize that the simulations are all based on the same exchange parameters obtained from the classical regime at $T\gg$\,\TN{}. We initially compare results from the lowest temperature ($T = 0.50$\TN{}) with minimal thermal fluctuations; see Figs.~\ref{spinwave}(a)-(b) and \ref{spinwave}(d). At such a low temperature, LLD produces nearly the same spin-wave spectrum as LSWT, with both methodologies converging on the same result as the temperature approaches zero \cite{supp}. Although the overall shape of the measured spectrum [Fig.~\ref{spinwave}(a)] resembles the LLD-generated spectrum [Fig.~\ref{spinwave}(b)], the match in terms of the dynamical susceptibility  $\chi''(|\mathbf{q}|,E)$ is not as precise as that found in the classical regime, implying the emergence of quantum effects not captured by the semi-classical LLD approach.

The primary discrepancy at $T = 0.50$\TN{} arises from LLD's underestimation of the overall magnon energy scale. This is attributed to the momentum-independent quantum renormalization factor $Z_{c} > 1$ of the magnon energy $\hbar \omega_{\mathbf{q}}$ in the N\'eel-ordered phase of the $S=1/2$ Heisenberg SqAF \cite{Sq_NLSWT1,Sq_NLSWT2,Sq_NLSWT3},
\begin{align}\label{eq:eigenE}
    \hbar \omega_{\mathbf{q}} =\, & 2J_{1}Z_{c} \sqrt{\left[1+\frac{j_{c}}{2}(1-\gamma_{\mathbf{q},c})\right]^2 - \gamma_{\mathbf{q},1}^2} \nonumber \\
    \equiv\, & 2\tilde{J}_{1}\sqrt{\left[1+\frac{j_{c}}{2}(1-\gamma_{\mathbf{q},c})\right]^2 - \gamma_{\mathbf{q},1}^2},    
\end{align}
where $j_{c} = |J_{c}|/J_{1}$ (with $J_1 > 0$ and $J_c < 0$), while $\gamma_{\mathbf{q},1} = \frac{1}{4} \sum_{\mathbf{\delta}_{1}} \mathrm{cos}(\mathbf{q}\cdot\mathbf{\delta}_{1})$ and 
$\gamma_{\mathbf{q},c} = \frac{1}{2} \sum_{\mathbf{\delta}_{c}} \mathrm{cos}(\mathbf{q}\cdot\mathbf{\delta}_{c})$ in terms of the intralayer and interlayer bond vectors $\mathbf{\delta}_{1}$ and $\mathbf{\delta}_{c}$. Typically, $Z_{c}$ is incorporated into $J_{1}$ by defining an effective coupling strength $\tilde{J}_{1}$, as in the second line of Eq.~(\ref{eq:eigenE}). This is because analyzing the spin-wave spectrum only reveals the value of $\tilde{J}_{1}=Z_{c}J_{1}$, leaving $J_{1}$ itself ambiguous unless the theoretical value of $Z_{c}$ is utilized \cite{La2CuO4}. However, by first analyzing the classical regime, $T\gg$\,\,\TN{} (Fig.~\ref{highT}), our methodology enables direct access to $J_{1}$ as the influence of $Z_{c}$ vanishes at sufficiently high temperatures. Consequently, the magnitude of $Z_{c}$ can be explicitly visualized through a comparison between the data [Fig.~\ref{spinwave}(a)] and the LLD simulations [Fig.~\ref{spinwave}(b)].

Moreover, assuming that $J_{1}$ is independent of temperature (which is mostly true unless a system undergoes a structural phase transition), $Z_{c}$ can be experimentally quantified by matching the energy scale of the LLD simulation to the observed spin-wave spectrum. For instance, as shown in Fig. \ref{spinwave}(d), $\tilde{J}_{1}\sim1.2J_{1}$ best describes the magnon energies measured at $T=$ 0.5\TN{}. Importantly, this value is consistent with the theoretical result $Z_{c} = 1.18(2)$ at zero temperature, predicted by spin-wave theory up to the order of 1/$S^{2}$ \cite{Zc_SWT} or the series expansion technique \cite{Zc_SE}. This validates the high precision of the exchange parameters derived from our refinement process in the classical regime (Fig.~\ref{highT}).

Another apparent discrepancy between the data and the LLD simulation result is the continuum scattering we observe above the one-magnon spectrum ($E>$\,1.8\,meV), as indicated by purple arrows in Figs.~\ref{spinwave}(a) and \ref{spinwave}(d). Notably, according to Fig. \ref{spinwave}(d), the spectral weight of this continuum scattering is approximately 17\% of that from the one-magnon modes. This continuum is likely a combination of two-magnon scattering that becomes most significant at $\mathbf{q} = (\pi, \pi)$ [$|\mathbf{Q}|$ = 0.71\,\AA\, in Fig.~\ref{spinwave}(a)] \cite{Sq_NLSWT1,Sq_NLSWT2,Sq_NLSWT3} and fractionalized quasi-particles that dominate the spectrum around $\mathbf{q} = (\pi, 0)$ [$|\mathbf{Q}|$ = 0.50 or 1.12\,\AA\, in Fig. \ref{spinwave}(a)] \cite{Christensen_2007,Mourigal_2015}. A comparison between the observed continuum signal and the theoretical dynamical structure factor of the two-magnon continuum is presented in the Supplemental Material \cite{supp}.

%%%%%%%%%%%%%%%%%%%%%%% Figure - Detailed T-dependence %%%%%%%%%%%%%%%%%%%%%%%%%%%%%%%%
\begin{figure}[t]
\includegraphics[width=0.8\columnwidth]{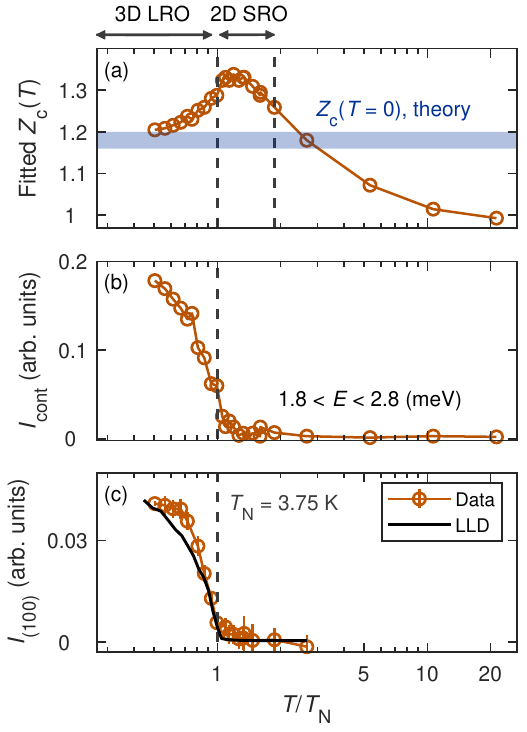} 
\caption{\label{T_dep} Temperature dependence of the quantum effects in the spin dynamics of ZVPO. (a) Temperature-dependent quantum renormalization factor $Z_{c}(T)$, obtained by dividing the optimal $\tilde{J}_{1}(T)$ at each temperature with $J_{1}$ = 0.695\,meV. Two arrows on the top indicate the ranges with 3D long-range order (LRO) and 2D short-range order (SRO) \cite{ZVPO_pnd}. (b) Temperature-dependent intensity of the quantum continuum ($I_{\mathrm{cont}}$), obtained by integrating the remnant intensities underestimated by LLD [orange solid lines in Figs.~\ref{spinwave}(e)-(g)] above 1.8 meV. (c) Temperature-dependent intensity of the $(100)$ magnetic Bragg peak. The data (open circles) are the same as the inset of Fig.~\ref{crystal}(d). Note that the reduced temperature $T$/\TN{} is plotted in a logarithmic scale for better presentation.}
\end{figure}
%%%%%%%%%%%%%%%%%%%%%%%%%%%%%%%%%%%%%%%%%%%%%%%%%%%%%%%%%%%%%%%%%%%%%%%%%%%%

Expanding the analysis to higher temperatures ($T>0.5$\TN{}) yields a complete temperature-dependent profile of $Z_{c}(T)$ and the continuum scattering. Figs.~\ref{spinwave}(e)-(g) show the measured energy dependence of $\chi''(|\mathbf{q}|,E)$, with $|\mathbf{q}|$ integrated from 0.55\,\AA$^{-1}$ to 1.5\,\AA$^{-1}$, and the corresponding LLD simulations with (orange solid lines) and without (blue dashed lines) the best-fitted $Z_{c}(T)$ for each temperature. We again emphasize that the calculations in Figs.~\ref{spinwave}(e)-(g) are all based on the same set of exchange parameters. The dashed lines are derived by applying an individual multiplicative scaling factor of $\chi''$ for each temperature, whereas the solid lines are obtained by fitting both the multiplicative scaling factor and the $Z_{c}$ parameter for each temperature. For selected temperatures, a detailed comparison of the energy- and momentum-resolved spectra is available in the Supplemental Material \cite{supp}. As long as the excitation energies are appropriately rescaled by the temperature-dependent but momentum-independent $Z_{c}(T)$, a good agreement is found for temperatures down to about 1.1\TN{} [Figs.~\ref{spinwave}(f)-(g)]. This illustrates LLD's capability to model the spin dynamics of the $S=1/2$ Heisenberg SqAF above \TN{}.

The resultant $Z_{c}(T)$ is shown in Fig. \ref{T_dep}(a). As expected, $Z_{c}(T)$ converges towards 1.18 for $T\ll T_{\mathrm{N}}$ and approaches 1 in the high-temperature regime ($T\gg T_{\mathrm{N}}$). Interestingly, however, the behavior of $Z_{c}(T)$ in the intermediate temperature range defies a simple, monotonic progression from $Z_{c}(T = 0) = 1.18$ to $Z_{c}(T\gg$\,\TN{}) = 1. Instead, $Z_{c}(T)$ gradually increases as $T$ approaches \TN{} and reaches its maximum between \TN{} and the temperature of maximum susceptibility, $T_{\mathrm{max}}\sim1.6$\TN{}, similar to the trend observed in the temperature-dependent magnetization [Fig.~\ref{crystal}(c)]. Nevertheless, we caution against over-interpreting this finding due to one limitation of our current LLD model: it relies on Boltzmann statistics to model thermal fluctuations, which likely results in an overestimation of thermal fluctuations around \TN{} \cite{rescale_Dahlbom}. Such an overestimation becomes apparent when comparing the measured and calculated intensities of the magnetic (100) peak [Fig.~\ref{T_dep}(c)], the latter of which was obtained from LLD simulations. As this would lead to the underestimation of excitation energies around $T=$\,\TN{}, the unusual behavior of $Z_{c}(T)$ near \TN{} may blend the actual quantum renormalization effect with adjustments for overestimated thermal fluctuations. Thus, accessing the true nature of $Z_{c}(T)$ around \TN{} first requires disentangling these two factors, which we leave as a future challenge. It can be potentially achieved by refining the current LLD approach to better follow the Bose-Einstein statistics, as suggested in Refs. \cite{color_noise,rescale_Dahlbom}. 

Analyzing the temperature dependence of the continuum scattering also reveals important insights. First of all, it is crucial to recognize that thermal fluctuations already induce a continuum-like signal above the one-magnon bands at sufficiently high temperatures (referred to as a ``thermal" continuum), as evident from the LLD simulation results shown in Figs.~\ref{spinwave}(e)-(f). Thus, the ``quantum" continuum, resulting from two-magnon scattering and fractionalized quasi-particles, must be distinguished by isolating the spectral weight not captured by the semi-classical LLD simulations. Indeed, LLD consistently underestimates $\chi''(|\mathbf{q}|,E)$ above the one-magnon band ($E > 1.8$\,meV) for $T$ below approximately 1.1\TN{} [see Figs.~\ref{spinwave}(d)-(f)]. Figure \ref{T_dep}(b) shows the integral of underestimated $\chi''(|\mathbf{q}|,E)$ above 1.8 meV at each temperature, exhibiting a similar trend to the magnetic Bragg peak intensity [Fig.~\ref{T_dep}(c)]. In other words, the emergence of the quantum continuum is marginal above \TN{}, implying its relevance to the formation of 3D long-range order. Conversely, the regime characterized by 2D short-range correlations (around $T_{\mathrm{max}}$) is still well described by semi-classical LLD simulations. For completeness, however, additional examination using $S=1/2$ SqAFs with significantly weaker $J_{c}$ (e.g., cuprates \cite{La2CuO4}) is required since the observed temperature dependence might have been affected by the non-negligible strength of $J_{c}$ in ZVPO ($5\sim7\,\%$ of $J_{1}$).

The comparative analysis of temperature dependence presented in this study demonstrates the proficiency of LLD in both accurately determining the spin Hamiltonian of quantum antiferromagnets and visualizing the quantum-to-classical crossover in their spin dynamics. Specifically, the remarkable agreement between the experimental data and the LLD simulations above \TN{} encourages the extension of this approach to more challenging $S=1/2$ systems with much stronger quantum fluctuations, in which the excitation spectra below \TN{} are expected to have larger deviations from predictions based on conventional spin-wave theory. Low-dimensional frustrated magnets, such as the $S=1/2$ triangular lattice antiferromagnet \cite{BCSO_INS, TLAF_nearQSL} or the $S=1/2$ honeycomb antiferromagnet proximate to the suggested Kitaev spin liquid phase \cite{RuCl3_sci}, stand out as promising examples, as also suggested in another recent theoretical study \cite{KSL_LLD}. 

\tred{The successful application of the LLD approach to powder-averaged data merits further discussion regarding both its implications and potential challenges. The extraction of optimal exchange parameters consistent with previous studies and the quantitative determination of $Z_{c}(T=0) = 1.19$, as previously calculated by theory, indicates that our LLD approach can be useful even for powder-averaged data. This capability not only broadens the scope of materials that can be explored experimentally but also allows the spectroscopic study of quantum antiferromagnets to proceed more rapidly. However, it is important to interpret the analysis results of powder-averaged spectra with caution. The analysis is typically challenged by the significant loss of information due to powder averaging, which can lead to considerable parameter uncertainty from the fitting or even potential non-unique solutions. Thus, a rigorous evaluation of the fitting results, such as the standard statistical analysis of parameters' uncertainties using reduced $\chi^{2}$ and exploring the $\chi^{2}$ map around the solution [Figs. \ref{highT}(h)--(i)], is essential. Also, the approach described here is not restricted to measurements of polycrystalline samples; it can also be readily applied to single-crystal studies. This will reveal deeper physical insights not evident in a powder-averaged spectrum, such as momentum-dependent magnon decay and/or renormalization. The LLD calculation protocol suggested in this work \cite{supp} can be directly applied to such future efforts.} Overall, our findings highlight the potential of LLD in exploring the dynamics of systems at the forefront of quantum magnetism research.

\begin{acknowledgments}
 We acknowledge Cristian Batista for the fruitful discussions about the LLD approach and its results. This research was supported by the U.S. Department of Energy, Office of Science, Basic Energy Sciences, Materials Science and Engineering Division. A portion of this research used resources at the Spallation Neutron Source, a DOE Office of Science User Facility operated by the Oak Ridge National Laboratory. Sample synthesis was supported by the Laboratory Directed Research and Development Program of Oak Ridge National Laboratory, managed by UT-Battelle, LLC, for the U. S. Department of Energy.
\end{acknowledgments}

% \bibliography{mainscript}% Produces the bibliography via BibTeX.

\providecommand{\noopsort}[1]{}\providecommand{\singleletter}[1]{#1}%
%

% \appendix

\clearpage
\onecolumngrid
% \section{A putative impurity signal in the heat capacity data of \yi{}}

\renewcommand{\thefigure}{A\arabic{figure}}
\renewcommand{\thetable}{A\Roman{table}}
\setcounter{figure}{0}
\setcounter{table}{0}

\end{document}